\documentclass[11pt]{article} 

\usepackage[utf8]{inputenc} 

\usepackage{geometry} 
\usepackage{graphicx} 

\usepackage{booktabs} 
\usepackage{array} 
\usepackage{paralist} 
\usepackage{verbatim} 
\usepackage{subfig} 
\usepackage{multicol}
\usepackage{sidecap}
\usepackage{wrapfig}

\usepackage{fancyhdr} 
\usepackage{sectsty}
\allsectionsfont{\sffamily\mdseries\upshape} 

\usepackage[nottoc,notlof,notlot]{tocbibind} 
\usepackage[titles,subfigure]{tocloft} 

\usepackage{indentfirst}

\title{\textbf{A neural network model that learns differences in diagnosis strategies among radiologists has an improved area under the curve for aneurysm status classification in magnetic resonance angiography image series}}
\author{}
\date{} 
\usepackage{amsmath}

\begin{document}
\LARGE
\begin{center}
\textbf{A neural network model that learns differences in diagnosis strategies among radiologists has an improved area under the curve for aneurysm status classification in magnetic resonance angiography image series}
\end{center}

\vspace{0.5cm}

\large
{\setlength\leftskip{1.0cm}
\noindent
Yasuhiko Tachibana$^{1}$, Masataka Nishimori$^{2}$, Naoyuki Kitamura$^{2}$, Kensuke Umehara$^{3}$, Junko Ota$^{3}$, Takayuki Obata$^{1}$, Tatsuya Higashi$^{4}$

}

\small
\vspace{0.5cm}
{\setlength\leftskip{1.0cm}
\noindent
1. Applied MRI Research, Department of Molecular Imaging and Theranostics, National Institute of Radiological Sciences, QST, Chiba, Japan

\noindent
2. MNES corporation, Hiroshima, Japan

\noindent
3. Medical Informatics Section, QST Hospital, QST, Chiba, Japan

\noindent
4. Department of Molecular Imaging and Theranostics, National Institute of Radiological Sciences, QST, Chiba, Japan

\vspace{0.5cm}
\noindent
Email: tachibana.yasuhiko@qst.go.jp

}

\subsection*{}

{\setlength\leftskip{2.0cm}
\noindent
\textbf{Abstract}

\noindent
\textbf{Purpose:}
To construct a neural network model that can learn the different diagnosing strategies of radiologists to better classify aneurysm status in magnetic resonance angiography images.

\noindent
\textbf{Methods:}
This retrospective study included 3423 time-of-flight brain magnetic resonance angiography image series (subjects: male 1843 [mean age, $50.2 \pm 11.7$ years], female 1580 [$50.8 \pm 11.3$ years]) recorded from November 2017 through January 2019. The images were read independently for aneurysm status by one of four board-certified radiologists, who were assisted by an established deep learning–based computer-assisted diagnosis (CAD) system. The constructed neural networks were trained to classify the aneurysm status of the zero to five aneurysm-suspicious areas suggested by the CAD system for each image series, and any additional aneurysm areas added by the radiologists, and this classification was compared with the judgment of the annotating radiologist. Image serieses were randomly allocated to training and testing data in an 8:2 ratio. The accuracy of the classification was compared by receiver operating characteristic analysis between the control model that accepted only image data as input and the proposed model that additionally accepted the input of who read the case. The DeLong test was used to compare areas under the curves ($P < 0.05$ was considered significant).

\noindent
\textbf{Results:}
The area under the curve was larger in the proposed model (0.845) than in the control model (0.793), and the difference was significant ($P < 0.0001$).

\noindent
\textbf{Conclusion:}
The proposed model improved classification accuracy by learning the diagnosis strategies of individual annotating radiologists.

}

\normalsize
\newpage

\begin{multicols}{2}

\section{Introduction}
Deep learning is an emerging technology that promises to move forward the field of diagnostic radiology. However, it is now understood that large amounts of data are required for successful training of deep learning models [1-3]. To address this issue, there are now many large clinical databases around the world, but compared with how much attention the need for greater amounts of data is attracting, the discussion regarding the homogeneity of data remains insufficient.

It is not easy to build a large clinical database containing only homogeneous data, and this task is made more difficult when subjective interpretations, such as those made by radiologists reading magnetic resonance (MR) angiography image series, are included as part of the data [4]. In such databases, as the amount of data increases, the number of individuals interpreting the data also increases, causing the homogeneity of the data to decrease.

Data inhomogeneity is a weak point of general supervised deep learning, because during the optimization process this type of deep learning assumes a unique output for each input [5]. In the case that an input can have several possible answers, such as when an image can be interpreted in several different ways, there is a risk that the trained model will output only a vague neutral answer instead of adequately converging to output a specific answer. This problem can be solved by performing training separately for each individual contributing data, but such an approach would decrease the amount of data available for each training. Another solution is to get a consensus for each judgment among the participating individual radiologists [6], but that is not realistic for large databases.

Here, we developed a neural network model for the radiological diagnosis context that can solve the problem of data inhomogeneity due to multiple radiologists contributing data by learning the differences among the radiologists’ diagnosis strategies during training. The model was designed for the task of classifying magnetic resonance (MR) angiography image patches as either aneurysm-positive or -negative. The purpose of this study was to evaluate whether the proposed strategy could improve the accuracy of classification, and to assess whether the strategy could help solve the data inhomogeneity problem.

\section{Materials and Methods}

This retrospective study was approved by our local ethics review board. Written informed consent was waived because the data used was acquired during daily routine practice.

\subsection{MRI scans, and data and annotator selection}
This study was performed using time-of-flight MR angiography image series of the brain recorded at our clinic from November 2017 through January 2019. Image series were obtained with a 1.5-T MRI system (Vantage Elan Zen, Canon, Tokyo) equipped with a 32-channel head coil. Parameters for the three-dimensional angiography scan were repetition time/echo time: 24/6.8 ms; flip angle: 20 degrees; field of view: 20 $\times$ 20 cm; matrix size: 512 $\times$ 512; bandwidth: 170 Hz/pixel; and scan time: approx. 6 min. The images used were the reconstructed two-dimensional images (slice thickness: 1.3 mm) used in clinical practice.

Image series were considered for inclusion as follows. First, series were selected in which the primary reading (before double-checking) was done by a board-certified radiologist (annotator) assisted by our clinic’s existing computer-assisted diagnosis system (CAD), which was developed on the basis of an established deep learning-based algorithm [7]. Next, the annotators for the selected series were ordered highest to lowest depending on how many series they had read, and annotators for whom the number of series read was more than one-third that of the annotator who had read the most series were selected. All image series interpreted by the selected annotators were included in the study, but those for which the study’s image processing could not be performed were excluded.

The frequency of an image series being judged as having one or more aneurysms ($F_{sub}$, \%) was compared among the annotators by using the Chi-Square test of independence (each possible pair was compared separately. $P <$ 0.01 was defined as significant, considering the multiple comparison).

\subsection{CAD system and the task given to the neural network}
The CAD system at our clinic suggests from zero to five aneurysm-suspicious areas in each image series; these areas are indicated as circles on the images. If the annotator agreed with the suggestion, they labeled it as “true positive (TP)”; if not, as “false positive (FP)”; and if they found an aneurysm that was not suggested, they added the new position and labeled it as “false negative (FN).” The frequency of judging a suggestion as true positive was compared among the annotators by using the Chi-Square test of independence ($P <$ 0.01 was defined as significant). The frequency of making a true positive judgment ($F_{CAD}$, [$F_{CAD} = TP/ (TP + FP)]$, \%) was also recorded for each annotator.

The task given to the neural network was the correct classification of TP, FP, and FN areas into the corresponding judgment—aneurysm-positive (TP and FN) or -negative (FP)—made by the annotators.

\subsection{Data sets}
Image series were divided into training and testing groups in an 8:2 ratio. This grouping was random but was made so that $F_{sub}$ was equal between the training and testing groups for each annotator.

To create input image data, a voxel of interest (VOI, 48 $\times$ 48 $\times$ 48 pixels) was designated around each target position. Then, maximum intensity projection was performed for each VOI in nine different directions to create a nine-channel image as the input image data (Figure 1); this subsampling was done in a manner similar to that in a previous report [6]. In the training group, subsampling was repeated up to 4000 times for each target with the VOI randomly shifted and rotated in three dimensions for each repetition. This repetition was done for data augmentation and to balance the sample sizes between combinations of annotator and aneurysm status.

\begin{SCfigure*}
\includegraphics[scale=0.5]{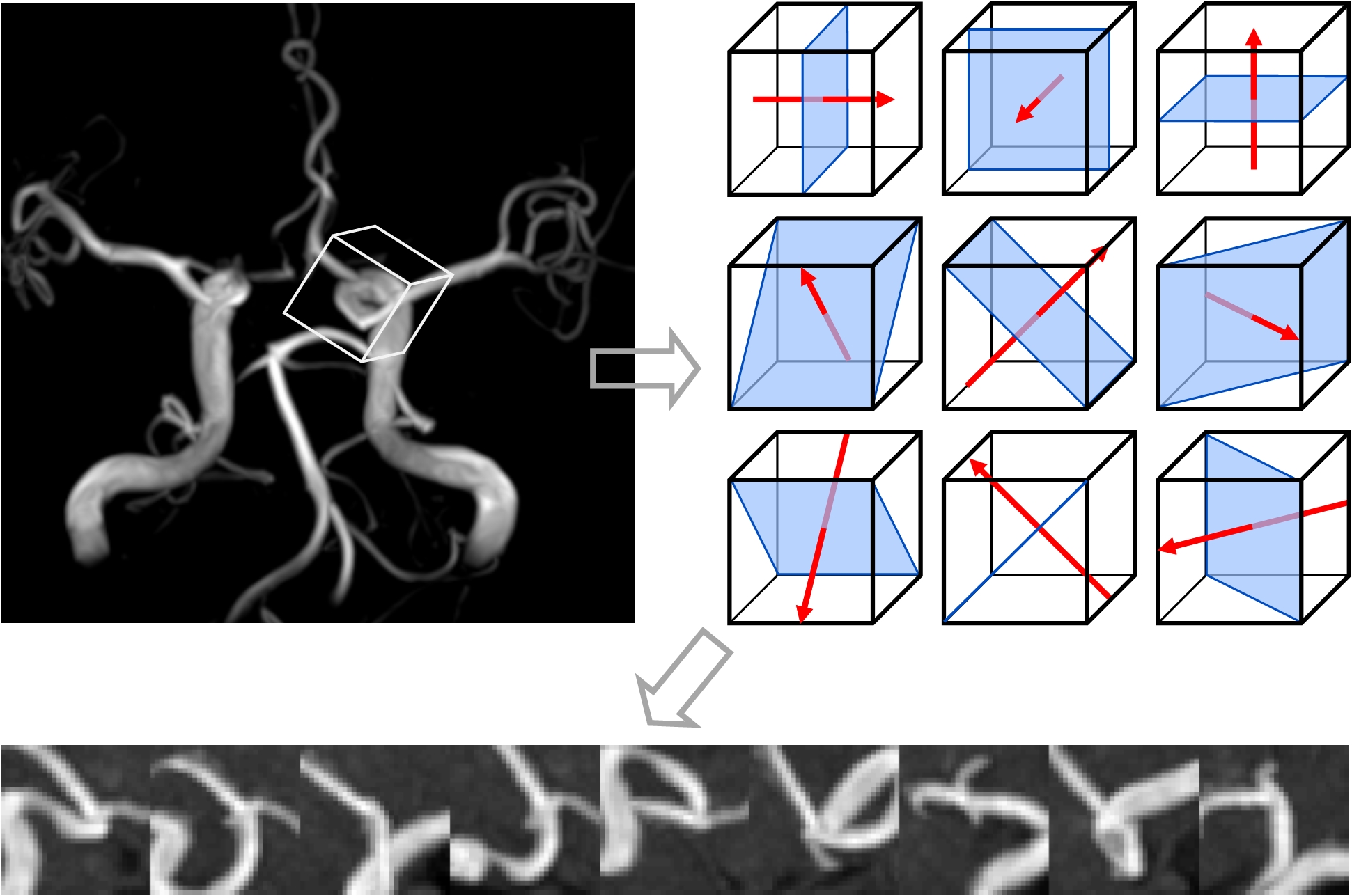}
\caption{\small\textbf{Data preprocessing to create input image data.} A voxel of interest (VOI, 48 $\times$ 48 $\times$ 48 pixels) was designated around the target position, and maximum intensity projection was performed in nine directions to create a nine-channel image for the VOI. (The nine channels are arranged side-by-side in this figure for visualization.) The process was repeated in the training group for data augmentation by shifting and rotating the VOI in three dimensions.}
\label{Figure 1}
\end{SCfigure*}

To create data sets from the training and testing groups, each nine-channel image was tagged with the aneurysm-positive or -negative judgment plus the following information: who the annotator was; whether the area was suggested by the CAD or not; and the recorded size of the aneurysm, if applicable. Each image series was assessed by only one annotator, without consensus.

\subsection{Model architecture}

\begin{SCfigure*}
\includegraphics[scale=0.8]{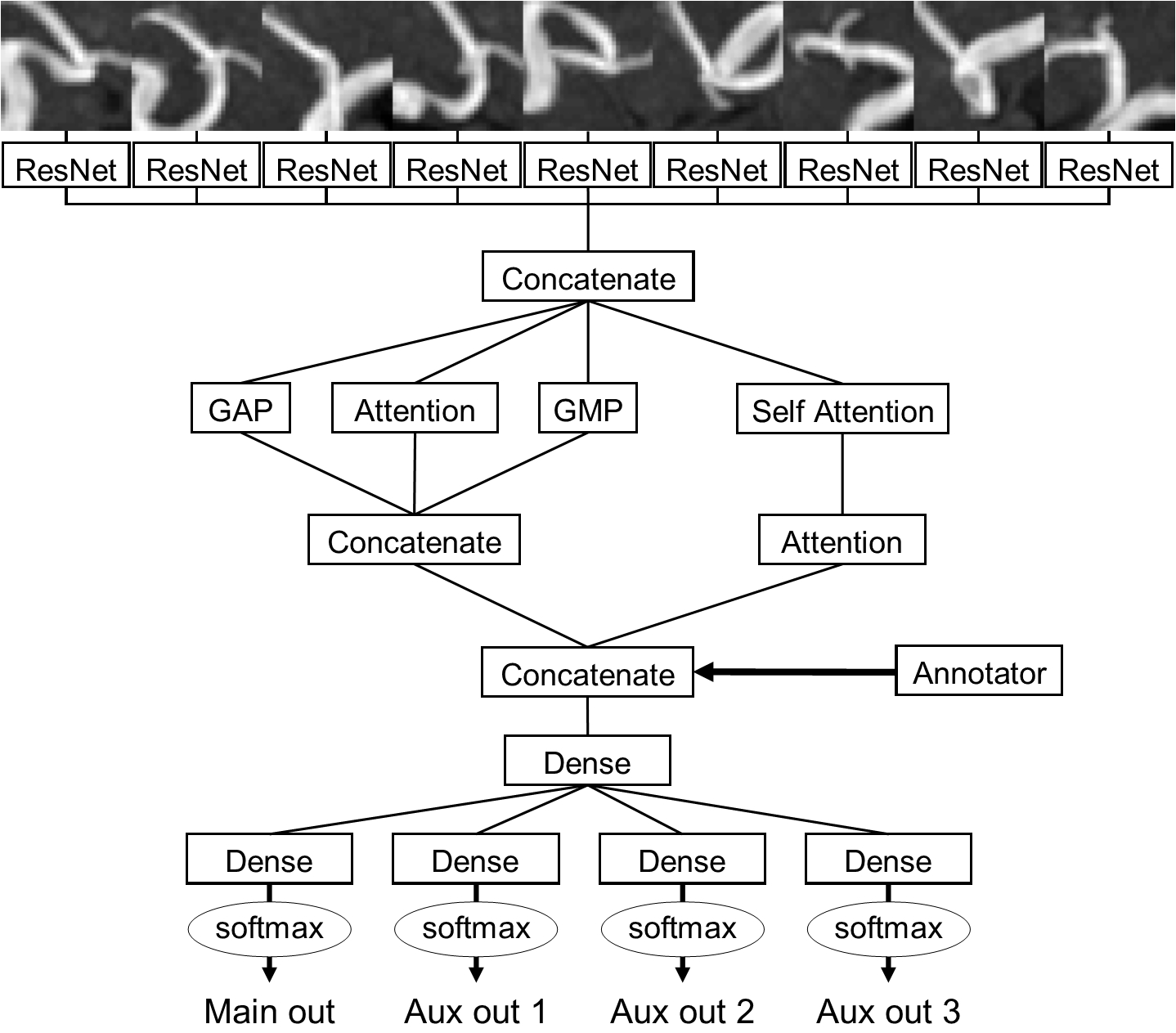}
\caption{\small\textbf{Network architecture of the proposed model.} Each channel of the input image was processed by a 20-layer ResNet structure and then concatenated after weighting. The information on who labeled the image (Annotator) was also merged. Finally, several densely connected layers were used to obtain the main output and three auxiliary outputs.
GAP: global average pooling layer; GMP: global max pooling layer; Dense: densely connected layer.
}
\label{Figure 2}
\end{SCfigure*}

The network architecture of the proposed model ($M_p$) is summarized in Figure 2. Input image data was first processed by a 20-layer ResNet structure [8]. Each channel of the input image data was processed in parallel, but the weights included in the ResNet structure were shared so that each image channel were equally processed. Then, after adjustment of their weight, the outputs from the ResNet structure were merged; a combination of several attention structures [9] was used to optimize the weighting before merging.

Then, the data of who the annotator was, which was translated to a one-hot vector, was also merged. Finally, the data was processed by one densely connected layer, and then by four parallel densely connected layers with softmax activation to obtain one main output and three auxiliary outputs. The main output was a length-two vector for the binary classification of aneurysm-positive or -negative. The auxiliary outputs were the following binary classifications: 1) judged as positive and recorded size of aneurysm was not zero, or not; 2) recorded size of aneurysm was $\geq$2 mm, or not; and 3) recorded size of aneurysm was $\leq$2 mm or smaller, or not. These outputs were added to stabilize the training.

To evaluate the usefulness of adding who the annotator was to the network, a control model ($M_0$) was prepared, which was exactly the same as $M_p$ except that it did not include the input of who the annotator was.

\subsection{Training, testing, and statistical comparisons between the models}
Training and testing were performed by using Keras [10] with the TensorFlow backend [11].

The models were trained by using the training data under the three-fold rule. The loss function was the categorical cross-entropy of the main and three auxiliary outputs merged after weighting 0.5, 0.1, 0.2, and 0.2, in this order. The RAdam algorithm [12] was used for optimization. Batch size and epoch were 32 and 200, respectively.

Models $M_0$ and $M_p$ were tested separately by using the test data. The data was inputted to the models trained in three patterns each (training was performed three-fold), and the square roots of the three main outputs were averaged according to an online discussion [13] to obtain the final output for the input.

The accuracy of classifying an input image to the corresponding annotator’s judgment was compared between $M_0$ and $M_p$ by receiver operating characteristic (ROC) analysis. Areas under the curve (AUC) were obtained for $M_0$ and $M_p$ by using all testing data at once (AUCall) and then separately for each annotator (AUC1 to AUC4). The DeLong test was used to statistically compare the AUCs between $M_0$ and $M_p$ ($P < 0.05$ was considered significant). MATLAB2019 (MathWorks Inc., Natic, USA) was used for all the statistical calculations.

\section{Results}
Figure 3 is a flowchart showing how the image series were selected. Of 14,141 image series recorded during the study period, 6369 were read with CAD assistance. Four radiologists (annotator 1–4), each with more than 7 years’ experience interpreting brain MR angiography, had read more than one-third the number of image series read by the annotator who had read the highest number of image series. Of the 6369 image series that had been read with CAD assistance, 3431 had been read by one of the four annotators. Eight image series were excluded because the target areas were around the periphery of the volume, so a VOI around the target could not be set. Finally, 3423 image series were included in the study (subjects: male 1843 [mean age, 50.2 $\pm$ 11.7 years], female 1580 [50.8 $\pm$ 11.3 years]).

\begin{figure*}
\begin{center}
\includegraphics[scale=0.7]{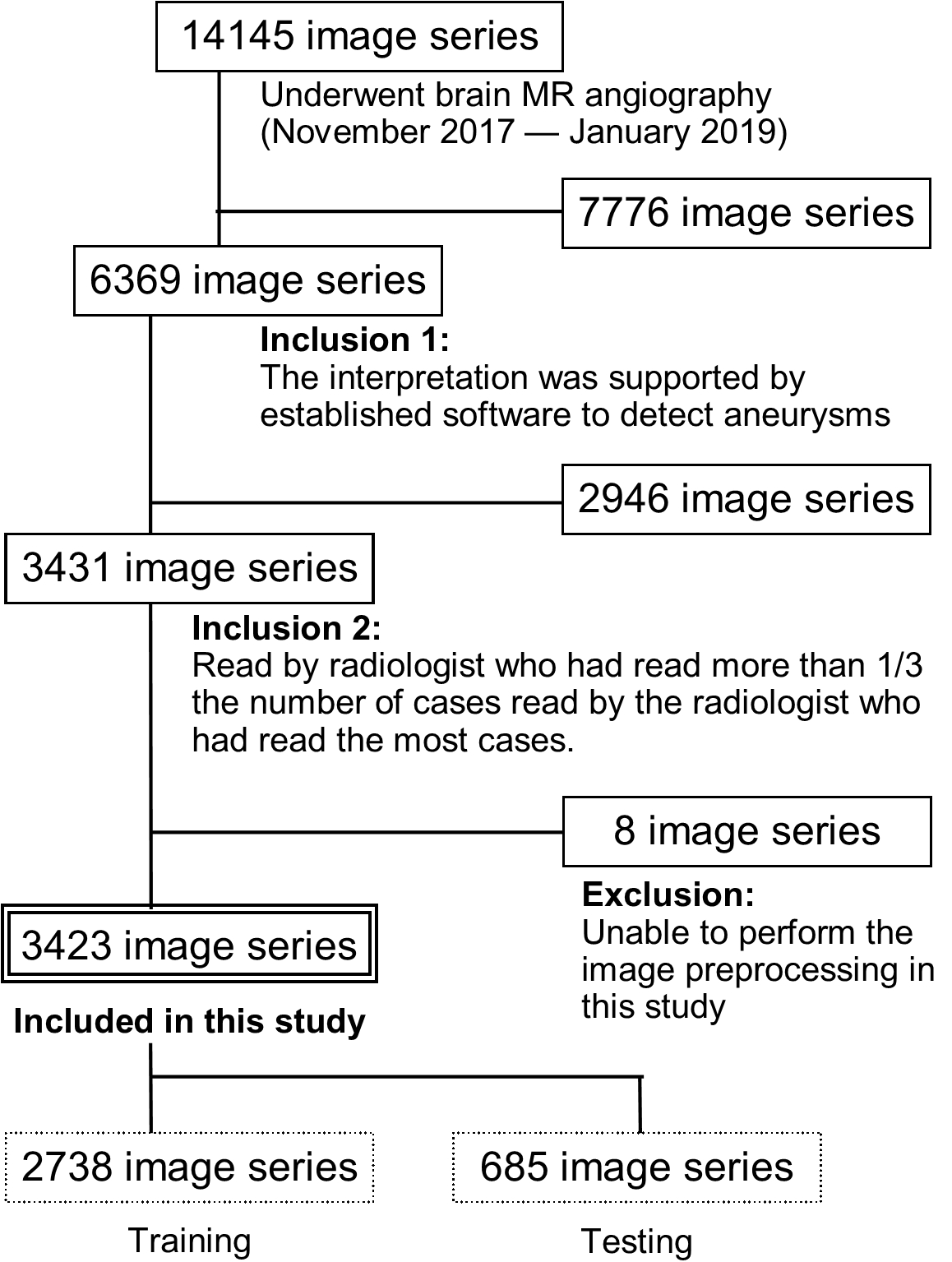}
\caption{\small\textbf{Flowchart for selecting image series.}}
\label{Figure 3}
\end{center}
\end{figure*}

Table 1 shows the judgments made by the annotators for the image series included in the training and testing groups. $F_{sub}$ increased in both groups in the order annotator 1, annotator 2, annotator 3, and annotator 4. $F_{sub}$ was comparable between annotators 2 and 3 (P = 0.32) but differed significantly between all other annotator pairs ($P <$ 0.0001).
The numbers of target areas assigned to the training and testing groups are shown in Table 2. $F_{CAD}$ increased in the same order as for $F_{sub}$. $F_{CAD}$ was again comparable between annotators 2 and 3 and differed significantly between all other annotator pairs ($P <$ 0.0001).

\begin{table*}
\caption{\textbf{Summary of the judgments made by the annotators}}
\label{Table 1}
\begin{center}
	\begin{tabular}{lcccccc}
		&\multicolumn{3}{c}{Training group} & \multicolumn{3}{c}{Testing group} \\
		&\multicolumn{2}{c}{Aneurysm status} &\multicolumn{1}{c}{$F_{sub}$} & \multicolumn{2}{c}{Aneurysm status} & \multicolumn{1}{c}{$F_{sub}$}\\ \cline{2-3} \cline{5-6}
				& Positive & Negative & (\%)	&Positive & Negative& (\%)	\\ \hline
	Annotator 1	& 17		& 625	& 2.6	& 4		& 156	& 2.5		\\
	Annotator 2	& 82		& 645	& 11.3	& 21		& 161	& 11.5		\\
	Annotator 3	& 122	& 838	& 12.7	& 31		& 209	& 12.9		\\
	Annotator 4	& 166	& 243	& 40.6	& 42		& 62		& 40.8		\\ \hline
	Total		& 387	& 2351	& 14.1	& 98		& 587	& 14.3		\\ \hline
		
	\end{tabular}
\end{center}
\small{$F_{sub}$: frequency of diagnosing one or more brain aneurysm [$F_{sub} = Positive/(Positive + Negative)$].}
\end{table*}

Data augmentation was not performed for the test data, but the data was made to represent the overall data set, for example, in terms of the number of cases read by each annotator and the number of aneurysm-positive cases for each annotator (Table 2). The training data set included a total of 580,916 data points, and the numbers of aneurysm-positive (TP + FN) and -negative (FP) tagged images were balanced for each annotator (Table 3).

The results of the ROC analysis are shown in Figure 4. The AUC was larger for $M_p$ than for $M_0$ in all of the comparisons. AUCall was 0.845 and 0.793 for $M_p$ and $M_0$, respectively. The difference between the AUCs for the two models was significant in the comparison for AUCall ($P <$ 0.0001) and also for AUC3 ($P =$ 0.0097).

Two representative outputs are shown in Figure 5. In addition to the predicted judgment for the corresponding annotator, predicted judgments for the annotators not reading the image series were also obtained by inputting the annotator information arbitrarily for each image.

\begin{table*}
\caption{\textbf{Summary of the numbers of target areas used for training and testing}}
\label{Table 2}
\begin{center}
	\begin{tabular}{lcccccccc}
		&\multicolumn{4}{c}{Training group} & \multicolumn{4}{c}{Testing group} \\
				& TP	& FP	& FN	&$F_{CAD}$ (\%)	& TP	& FP	&FN		& $F_{CAD}$ (\%)	\\ \hline
	Annotator 1	& 17		& 1469	& 0		& 1.1			& 4		& 358	& 0		& 1.1 			\\
	Annotator 2	& 84		& 1584	& 4		& 5.0			& 22		& 427	& 0		& 4.9			\\
	Annotator 3	& 138	& 2072	& 1		& 6.2			& 33		& 549	& 0		& 5.7			\\
	Annotator 4	& 146	& 794	& 57		& 15.5			& 39		& 198	& 6		& 16.5			\\ \hline
	Total		& 385	& 5919	& 62		& 6.1			& 98		& 1532	& 6		& 6.0			\\ \hline
	\end{tabular}
\end{center}
\small{TP: true positive, target suggested as an aneurysm by the computer-assisted diagnosis (CAD) system and agreed on by the annotator; FP: false positive, target suggested as an aneurysm by the CAD system but not agreed on by the annotator; FN; false negative, target not suggested by the CAD system but added by the annotator as an aneurysm; $F_{CAD}$: frequency of the annotator agreeing with the CAD system suggestion [$F_{CAD} = TP/(TP + FP)$]}
\end{table*}

\begin{table*}
\caption{\textbf{Summary of the data in the training data set after data augmentation}}
\label{Table 3}
\begin{center}
	\begin{tabular}{lccc}
				& Aneurysm-positive (TP+FN)	& Aneurysm-negative (FP)		& Total		\\ \hline
	Annotator 1	& 67387						& 67387						& 132774		\\
	Annotator 2	& 69548						& 69548						& 139096		\\
	Annotator 3	& 76123						& 76123						& 152246		\\
	Annotator 4	& 78400						& 78400						& 156800		\\ \hline
	Total		& 290458						& 290458						& 580916		\\ \hline
	\end{tabular}
\end{center}
\small{TP: true positive, target suggested as an aneurysm by the computer-assisted diagnosis (CAD) system and agreed on by the annotator; FP: false positive, target suggested as an aneurysm by the CAD system but not agreed on by the annotator}
\end{table*}

\begin{figure*}
\begin{center}
\includegraphics[scale=0.6]{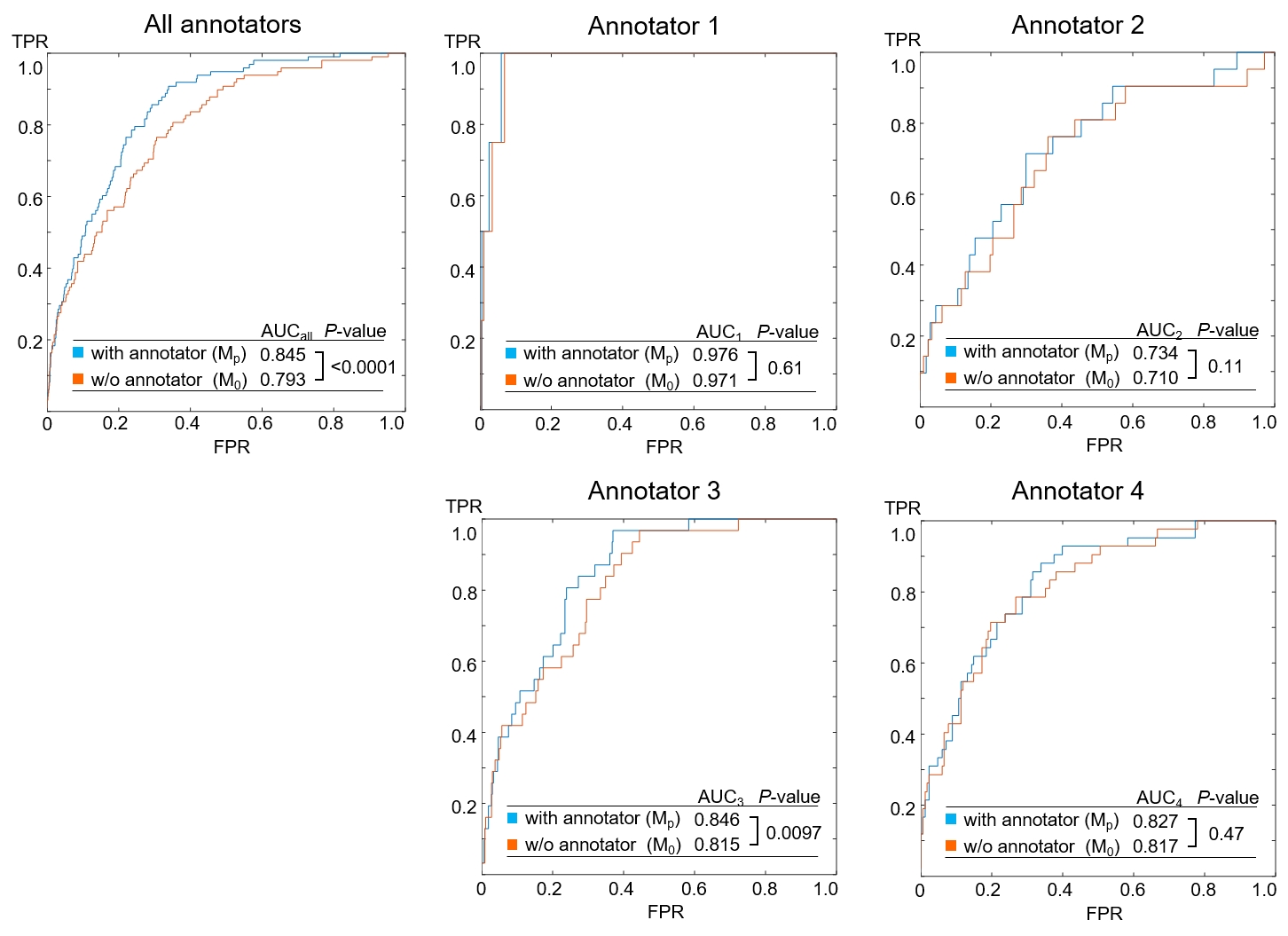}
\caption{\small\textbf{Receiver operating characteristics curves for models with and without the annotator information ($M_p$ and $M_0$, respectively).} The area under the curve (AUC) was larger for $M_p$ than for $M_0$ in all comparisons. The difference between the models was significant for AUCall ($P <$ 0.0001, DeLong test) and AUC3 (P = 0.0097). TPR: true positive ratio; FPR: false positive ratio; AUCall: AUC obtained from the whole test data; AUC1 to AUC4: AUCs obtained separately for annotator 1 to 4}
\label{Figure 4}
\end{center}
\end{figure*}

\begin{figure*}
\begin{center}
\includegraphics[scale=0.9]{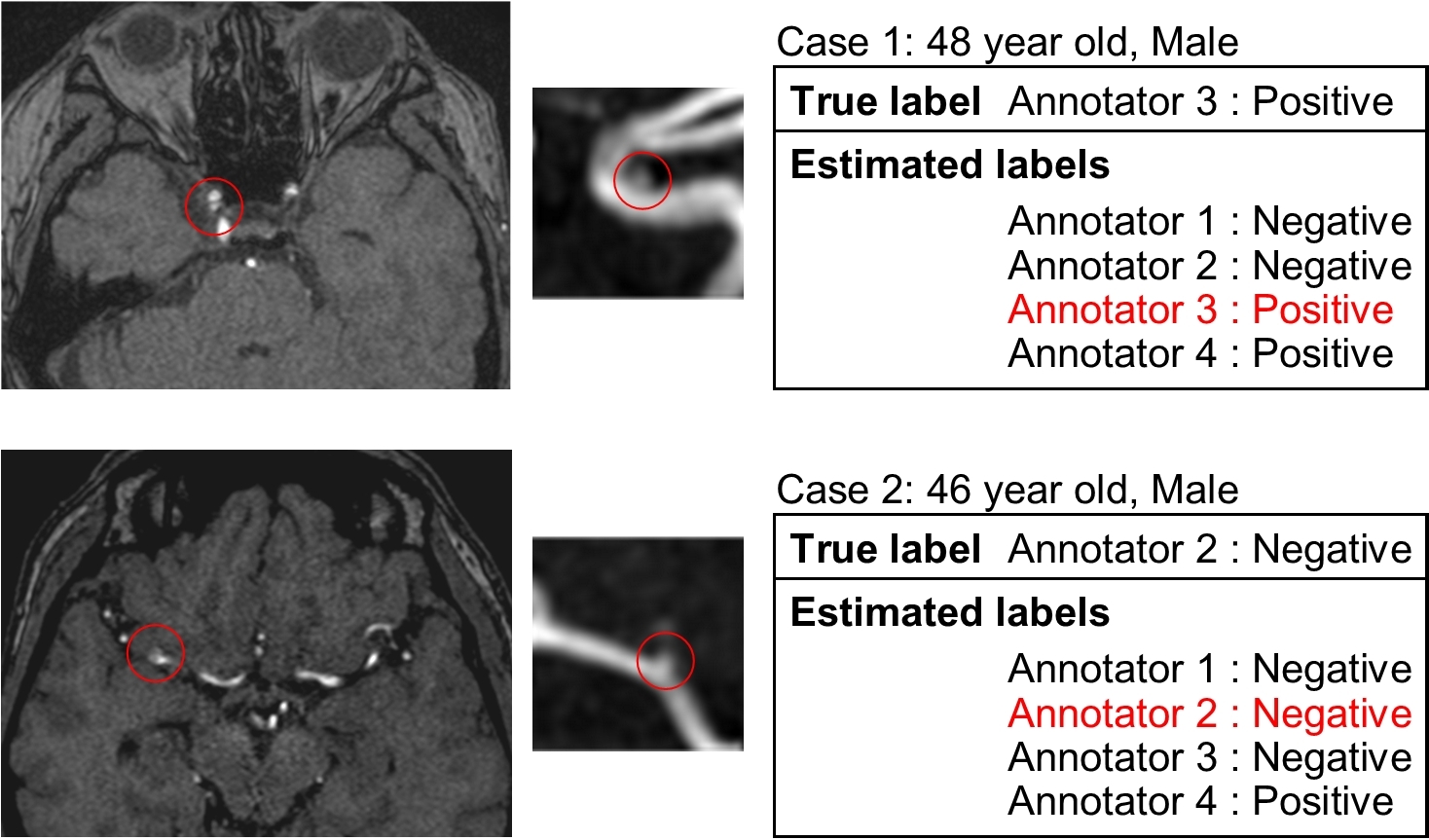}
\caption{\small\textbf{Representative outputs.} Target areas (red circles) are indicated on each magnetic-resonance angiography image (left: axial slice including the target, right: representative maximum-intensity-projection image for the same target). The true label is that assigned by the indicated annotator, whereas the estimated labels are predictions made by the trained model by changing the annotator information arbitrarily for the input image. The estimated label for the annotator who read the image (highlighted in red) was used for analysis.}
\label{Figure 5}
\end{center}
\end{figure*}

\section{Discussion}
Here, we evaluated whether incorporating information on the annotator of an MR angiography image series into a network model would allow the model to consider differences in diagnosis strategy among the annotators and improve classification accuracy. Two models were constructed, $M_p$ and $M_0$, which differed only by whether the input included who the annotator was, and we found that AUCall was significantly higher in $M_p$ than in $M_0$, suggesting that our strategy reduced the negative effect of inhomogeneous data interpretation by allowing the model to consider the individual diagnosis pattern of each annotator.

This study was built on two assumptions. The first is that there are differences in annotators’ diagnosis strategies that decreased training performance in the general model (i.e., $M_0$). The second is that those different diagnosis strategies are not arbitrary but instead are based on a certain logical paradigm each annotator has cultivated through their expertise, such that it can be learned by the optimized neural network.

The first assumption, that the diagnosis strategies differed among the annotators, is supported by the distribution of the judgments made by the annotators. That is, when the annotators were arranged in numerical order for $F_{sub}$ (frequency of diagnosing an aneurysm) and $F_{CAD}$ (frequency of agreeing with the CAD suggestion), the same order was observed and the differences between each annotator pair were all significant, except for between annotator 2 and 3. Because it can be assumed that both the overall data set and the CAD decision data were homogenous, these differences most likely reflect the differences in the diagnosis strategies of the annotators. It should be noted here that, because the annotator judgments were based mostly on suggestions provided by the CAD system, the effect of simple human error, such as overlooking a lesion, was likely negligible.

The second assumption, that individual diagnosis strategies are based on a unique, learnable, logical diagnosing paradigm, is supported by the result that AUCall for $M_p$ was significantly improved compared with that for $M_0$. If the diagnosis strategy was inconsistent (or somewhat random) for each annotator, and the differences in $F_{sub}$ and $F_{CAD}$ among the annotators were only reflecting the tendency for a positive judgment, this result would not have been achieved, because the proportions of aneurysm-positive and -negative image areas in the training data were adjusted to be equal for each annotator.

The AUC comparisons for each annotator indicated a significant difference between the two models only for annotator 3. However, this result may not discount using the proposed method. First, for annotator 1, the AUC for $M_0$ was already high (0.97). Supposing, given that annotator 1 had the lowest $F_{sub}$ and $F_{CAD}$, that the sensitivity of annotator 1 was the lowest among the annotators and that the specificity was the highest, his or her positive judgment might have been given only in cases where the other annotators also gave a positive judgment. If this is true, then the difference in judging strategies among the annotators is likely not an issue, even in $M_0$. Second, for all annotators, a situation in which the AUCs of $M_0$ and $M_p$ are similar can occur when the converged outputs of $M_0$ happen to not be too far from $M_p$ for that annotator. Therefore, given that there was a significant improvement in AUCall, the fact that AUC1 to AUC4 were all larger for $M_p$ than for $M_0$ is more important than whether they were statistically significant.

This study did not compare the accuracy of the two models on the basis of a ground truth diagnosis; instead, it examined only whether the model output matched each annotator’s judgment. However, because it is difficult to define a ground truth for a positive or negative classification (especially for small aneurysms), it might be useful in clinical practice to have a comprehensive view of how much opinions agree or disagree when multiple clinicians with various diagnosis strategies make a diagnosis for the same case. This can be achieved with the proposed network, because it accepts arbitrary input of who the annotator is and so it can output the estimated judgment for all annotators (Figure 5); that is, a “virtual conference” can be held.

There were several limitations in this study. First, only the data of four radiologists was considered. Some radiologists were not included because they had read too few cases. This selection stage was important to ensure there were sufficient test cases for each annotator. Second, we did not compare the accuracy of the developed model with that of the current CAD system. General comparisons based on ROC or accuracy and specificity were not available because the “true negative” judgment of the CAD was invisible. Furthermore, because the training and testing data was taken from the same clinical database, whereas the CAD system was trained by images taken from a different database, it was not appropriate to use our testing data to compare $M_p$ and CAD; another unrelated test data set would be required for such a purpose. However, the high AUC (0.845) achieved by $M_p$ suggests a potential advantage of $M_p$ over the present CAD system.

To conclude, the proposed network model could classify images appropriately to each annotator’s judgment by considering differences among diagnosis strategies. This approach could be useful when building a diagnostic support tool, like the CAD system used here, from a clinical database that includes interpretations made by multiple clinicians. Alternatively, the proposed approach could be suitable for other deep learning tasks where the data set is expected to contain a certain consistent trend bias; for example, in a multicenter study based on images acquired by using different MRI machines. When databases become larger and more clinicians are participating in data collection, we are sure that the merits of this approach will become more apparent.

\section*{Acknowledgement}

This research was supported by a Grant-in-Aid for Scientific Research (Kakenhi 17K10385) from the Japan Society for the Promotion of Science (JSPS) and Japanese Government.
The authors appreciate the assistance of Masaaki Chiku, Toshiyuki Morito, Hiroko Kamada, and Etsuko Mitsui during the study.

\end{multicols}
\end{document}